\documentclass[12pt,preprint]{emulateapj}
\usepackage{graphicx}
\usepackage{subfigure}

\shorttitle{Contemporaneous Imaging of $\sigma$~Geminorum}
\shortauthors{Roettenbacher et al.}
\slugcomment{Accepted to ApJ}

\begin{document}

\title{Contemporaneous Imaging Comparisons of the Spotted Giant $\sigma$~Geminorum Using Interferometric, Spectroscopic, and Photometric Data}
\author{
  Rachael M.\ Roettenbacher$^{1,2}$, John D.\ Monnier$^2$, Heidi Korhonen$^{3,4}$, Robert~O.~Harmon$^5$, Fabien Baron$^{2,6}$,  
  Thomas Hackman$^7$, Gregory W.\ Henry$^8$, Gail~H.~Schaefer$^9$, Klaus G. Strassmeier$^{10}$, Michael~Weber$^{10}$, 
  and Theo~A.~ten~Brummelaar$^9$
  }
    \affil{$^1$Department of Astronomy, Stockholm University, SE-106 91 Stockholm, Sweden \\
$^2$Department of Astronomy, University of Michigan, Ann Arbor, MI 48109, USA \\  
     $^3$Dark Cosmology Centre, Niels Bohr Institute, University of Copenhagen, DK-2100 Copenhagen \O, Denmark \\
     $^4$Finnish Centre for Astronomy with ESO (FINCA), University of Turku, FI-21500 Piikki\"o, Finland \\
  $^5$Department of Physics and Astronomy, Ohio Wesleyan University, Delaware, OH 43015, USA \\ 
  $^6$Department of Physics and Astronomy, Georgia State University, Atlanta, GA 30303, USA\\
  $^7$Department of Physics, PO Box 64, FI-00014, University of Helsinki,  Helsinki, Finland\\
  $^8$Center of Excellence in Information Systems, Tennessee State University, Nashville, TN 37209, USA \\
$^9$Center for High Angular Resolution Astronomy, Georgia State University, Mount Wilson, CA 91023, USA \\
$^{10}$Leibniz-Institute for Astrophysics Potsdam (AIP), D-14482 Potsdam, Germany \\
}
\email{rachael@astro.su.se}

\begin{abstract}

Nearby, active stars with relatively rapid rotation and large starspot structures offer the opportunity to compare interferometric, spectroscopic, and photometric imaging techniques.  In this paper, we image a spotted star with three different  methods for the first time.  The giant primary star of the RS Canum Venaticorum binary $\sigma$~Geminorum ($\sigma$~Gem) was imaged for two epochs of interferometric, high-resolution spectroscopic, and photometric observations.  
The light curves from the reconstructions show good agreement with the observed light curves, supported by the longitudinally-consistent spot features on the different maps.  However, there is strong disagreement in the spot latitudes across the methods.  

\end{abstract}



\section{Introduction}

The magnetic fields of cool stars can be strong enough to suppress convection in the outer layers.  These regions of stifled convection---starspots---are dark against the bright photosphere \citep[][and references therein]{ber05,str09}.  A particular class of active stars known to exhibit large starspots are RS Canum Venaticorum (RS CVn) variables.  These are customarily binary systems with an evolved giant or subgiant primary component that is often tidally-locked in a close orbit (typical orbital periods are on the order of $1-3$ weeks) with a less-evolved, usually main-sequence companion.  RS CVn primary stars are known to be active through rotational modulations in photometric and Ca H \& K observations \citep{hal76}.  Because of their activity RS CVn primary stars are often targets of imaging studies to map the stellar magnetic fields \citep[e.g.,][]{ros15} and the surface features \citep[e.g.,][]{kov15}.  Spotted stellar surface maps have made use of interferometric, spectroscopic, and/or photometric data sets.  

Interferometric aperture synthesis imaging is a direct imaging method that offers an independent estimate of stellar parameters and does not require \emph{a priori} knowledge of the stellar surface features.  This method allows for the stellar surface to be mapped as it appears on the sky, revealing, for example, stellar inclination and position angle  \citep[e.g.,][]{mon07}. 
The method has previously been used to image rapidly-rotating stars, expanding ejecta from novae, and starspots \citep{mon07,sch14,roe16}.  Interferometry combines light from two or more telescopes to obtain resolutions not accessible by individual telescopes, which are limited by mirror size.  An interferometric array of telescopes mimics a single telescope with a mirror diameter equal to the longest baseline (distance between two telescopes).
The angular resolution of an interferometric array is limited only by the length of the longest interferometric baseline in the array of telescopes.  Uniquely, interferometry is the only technique able to distinguish temperatures across the stellar surface in latitude and longitude, which has been used to map stellar surfaces in order to measure gravity darkening on, for example, rapidly-rotating stars \citep[e.g.,][]{zha09,che11}.

High-resolution, high signal-to-noise  spectroscopic data of a star at different rotational phases can be used to create  maps of stellar surfaces through Doppler imaging \citep[e.g.,][]{vog87,ric89,pis91}. 
This technique is used for detailed mappings of the stellar surface structures: temperature spots, chemical inhomogeneities, and surface magnetic fields (with spectropolarimetric observations).  In order to produce the most accurate surface reconstructions, Doppler imaging requires detailed estimates of stellar parameters to model the stellar absorption lines.  
As the star rotates, the effects of the starspots are observed as rotationally-modulated distortions in the absorption line profiles. These distortions are produced by the inhomogeneities of the stellar surface.  By tracking the changes in the absorption lines as the star rotates, the surface of the star can be reconstructed, revealing information about spot longitude and latitude.  However, Doppler imaging cannot always reliably reconstruct the hemisphere of the spot, particularly in the hemisphere of the hidden pole.  In Doppler imaging, the star's mean temperature profile as a function of latitude does not generate a time-variable signature, which means that the profile is sensitive to the assumptions made in absorption line profile modeling.  In particular, there is a degeneracy between the microturbulence and mean temperature. 

Photometric light-curve inversion algorithms use one or more light curves to reconstruct the rotationally-modulated features of a stellar surface \citep[e.g.,][]{har00,sav08}.  These techniques allow for accurate determinations of the longitude of surface features and can be applied to a wide variety of rotating stars.  To reduce degeneracies, light-curve inversion requires knowledge of some stellar parameters (e.g., spot and photosphere temperatures and stellar inclination).  A major drawback of  light-curve inversion  is its limitation in latitude determination, which can only be partially rectified by multi-bandpass observations.  Light-curve inversion is also unable to make distinctions of surface temperature.

In this paper, we compare these three imaging methods on two epochs of contemporaneous data of an RS CVn binary system that is particularly well-suited for imaging:  the close, tidally-locked, giant primary of $\sigma$~Geminorum ($\sigma$~Gem, HD 62044). $\sigma$~Gem is a spotted RS CVn binary \citep[e.g.,][]{hen95} with known orbital and stellar parameters \citep{roe15}.  The system consists of a resolved, inflated primary star (limb-darkened diameter $\theta_\mathrm{LD} = 2.417 \pm 0.007$, $M_1 = 1.28 \pm 0.07 \ M_\odot$, $R_1 = 10.1 \pm 0.4 \ R_\odot$) and an unresolved, main-sequence companion ($M_2 = 0.73 \pm 0.03 \ M_\odot$) in a circular orbit with semi-major axis $a = 4.63 \pm 0.04$ mas, inclination $i = 107.7 \pm 0.8^\circ$, orbital period $P_\mathrm{orb} = 19.6027 \pm 0.0005$ days, and distance $d = 38.8 \pm 0.6$ pc   \citep{roe15}. 
$\sigma$~Gem has been a target of Doppler imaging \citep{hat93, kov01, kov15} and long-term ground-based photometric monitoring for studies of spot activity \citep{hen95, ber98b, kaj14}.  

Here, the three imaging techniques---interferometric aperture synthesis, Doppler, and light-curve inversion imaging---are compared for two observational epochs $\sigma$~Gem.
In one epoch of observation (2011), $\sigma$~Gem has a simple starspot structure with two close, strong starspots, and in the other epoch (2012), $\sigma$~Gem has a more complex, global starspot network.  
By comparing the imaging techniques in these two different cases, we highlight the advantages and disadvantages of each method. 
We discuss our observational data sets in Section 2.  In Section 3, we briefly describe the imaging algorithms used and show their resultant images.  In Section 4, we discuss the images through comparison.  In Section 5, we report our conclusions on the starspots of $\sigma$ Gem and our comparative imaging.  


\section{Observations}

Simultaneous data sets of $\sigma$~Gem were obtained for the first time at a variety of facilities during two epochs for comparative imaging of the spotted star.  In order to obtain the most complete phase coverage possible, these observations span more than one rotation in some cases.  While the features on the surface of $\sigma$~Gem change over time, the starspots do not evolve rapidly enough for phase-folding over a small number of rotation periods to show significant evolution of stellar activity.  \citet{hus02}, for example, showed that starspots on evolved stars like $\sigma$~Gem evolve more slowly than those of young stars.

\subsection{Interferometry}

Interferometric data of $\sigma$~Gem were obtained with Georgia State University's Center for High Angular Resolution Astronomy (CHARA) Array at Mount Wilson Observatory, USA.  The CHARA Array consists of six 1-m class telescopes arranged in a non-redundant Y-shaped array with baselines ranging from 34 to 331 m \citep{ten05}.  Using the Michigan InfraRed Combiner \citep[MIRC;][]{mon04} with all six CHARA telescopes, we observed $\sigma$~Gem in the $H$-band (eight channels across $1.5-1.8$ $\mu$m for $\lambda/\Delta\lambda \sim 40$).  Our observations occurred on UT 2011 November 9 and December 7, 8, 9; 2012 November 7, 8, 21, 22, 24, 25 and December 4, 5, 7, 8 (see Table \ref{mircobs} for the telescopes used and the numbers of squared visibilities and closure phase data points for each night).  

We reduced and calibrated these data with the standard MIRC pipeline \citep{mon12}.  At least one calibration star was used each night (see Table \ref{cal}).  The data products resultant from the MIRC pipeline are observables including visibility, closure phases, and triple amplitudes (see sample observations in Figures \ref{2012Vis}--\ref{2012T3}, located in Appendix A).
On five nights of observation (UT 2011 December 8; 2012 November 7, 8, 24, and 25), we detected the companion star and combined these interferometric detections with radial velocity observations to determine the orbital and stellar parameters \citep{roe15}.  

\begin{deluxetable*}{l c c c}
\tabletypesize{\scriptsize}
\tablecaption{MIRC Observations of $\sigma$~Geminorum}
\tablewidth{0pt}
\tablehead{
\colhead{UT Date} & \colhead{Telescopes Used} & \colhead{Number of} & \colhead{Number of}\\
 &  & \colhead{$V^2$} & \colhead{Closure Phases}
}
\startdata
2011 Nov 9 & S1 - S2 - E1 - E2 - W2 & 48 & 32\\
2011 Dec 7 & S1 - S2 - E1 - E2 - W2 & 62 & 40\\
2011 Dec 8 & S1 - S2 - E1 - E2 - W1 - W2 & 399 & 432\\
2011 Dec 9 & S1 - S2 - E1 - E2 - W2 & 71 & 80\\
2012 Nov 7 & S1 - S2 - E1 - E2 - W1 - W2 & 799 & 1039\\
2012 Nov 8 & S1 - S2 - E1 - E2 - W1 - W2 & 379 & 480\\
2012 Nov 21 & S1 - S2 - E1 - E2 - W2 & 80 & 79\\
2012 Nov 22 & S1 - S2 - E1 - E2 - W2 & 170 & 133\\
2012 Nov 24 & S1 - S2 - E1 - E2 - W1 - W2 & 287 & 400\\
2012 Nov 25 & S1 - S2 - E1 - E2 - W1 - W2 & 480 & 640\\
2012 Dec 4 & S1 - S2 - E1 - E2 - W2 & 85 & 64\\
2012 Dec 5 & S1 - S2 - E1 - E2 - W2 & 154 & 129\\
2012 Dec 7 & S1 - S2 - E1 - E2 - W1 - W2 & 214 & 256\\
2012 Dec 8 & S1 - S2 - E1 - E2 - W1 - W2 & 213 & 191

\enddata
\label{mircobs}
\end{deluxetable*}

\begin{deluxetable*}{l c c c}
\tabletypesize{\scriptsize}
\tablecaption{Calibrators for $\sigma$~Geminorum}
\tablewidth{0pt}
\tablehead{
\colhead{Calibrator Name} & \colhead{Calibrator Size (mas)} & \colhead{Source} & \colhead{UT Date of Observation}
}
\startdata
HD 24398 ($\zeta$ Per) & $0.70 \pm 0.03$ & \cite{bar78} & 2011 November 9, December 7\\
HD 37329 & $0.71 \pm   0.05$ & \cite{bon06} & 2011 December 7, 8; 2012 November 8\\
HD 50019 ($\theta$ Gem) &  $0.81 \pm  0.06$ & \cite{bon06} &  2011 November 9, December 8; \\
& & & 2012 November 7, 8, 24, 25, December 4, 5, 7, 8 \\
HD 52711 & $0.62 \pm 0.05$ & \cite{bon06} & 2011 December 9\\
HD 63138 & $0.65 \pm  0.04$ & MIRC calibration & 2011 December 8, 9; 2012 November 8\\
HD 69897 ($\chi$ Cnc) & $0.73 \pm 0.05$ & \cite{bon06} & 2011 November 9, December 7, 8; \\
& & & 2012 November 7, 21, 22, 24, 25, December 4, 5, 7, 8
\enddata
\label{cal}
\end{deluxetable*}

\subsection{Spectroscopy}

Optical high-resolution spectra of $\sigma$~Gem were obtained at fifteen epochs between 2011 October 24 and November 9 using the fiber-fed STELLA Echelle Spectrograph (SES) at the robotic 1.2-m STELLA-II telescope at Izana Observatory, Tenerife, Spain \citep{str04}. SES covers wavelengths from  388 nm to 882 nm in a single exposure with spectral resolution ($\lambda/\Delta\lambda$) of 55,000 using two pixel sampling. For all spectra the exposure time was set to 900 seconds resulting in signal-to-noise ratio (S/N) per resolution element between 107 and 201 at 642.5 nm. The observing scheme also included nightly flat-fields, bias frames, and Thorium-Argon wavelength calibration exposures. The data were reduced using a dedicated SES pipeline \citep{web08}. A detailed observing log with the observing dates, stellar rotational phases, calculated orbital velocity, and S/N per resolution element is given in Table~\ref{spectra_log}.

$\sigma$~Gem was also observed on seven epochs between 2011 November 5 and November 28, and on eight epochs between 2012 November 8 and December 14 using the UV-Visual Echelle Spectrograph \citep[UVES;][]{dek00} mounted on the 8-m Kueyen telescope of the Very Large Telescope (VLT; Paranal Observatory, Chile). For these observations the red arm of the spectrograph was used with the imageslicer \#3 in the standard wavelength setting of 580 nm. This instrument setup gives a spectral resolution of 110,000 with two pixel sampling, and a wavelength coverage of 500--700 nm. The exposure time of each observation was 12 seconds yielding S/N per resolution element between 348 and 611 (at 642.5 nm). The standard UVES calibration plan together with the UVES pipeline were used for the data reduction. A summary of the UVES observations is also given in Table~\ref{spectra_log}.

\begin{deluxetable*}{l c c c}
\tabletypesize{\scriptsize}
\tablecaption{Spectroscopic Observing Log}
\tablewidth{0pt}
\tablehead{
\colhead{Modified Julian Date} & \colhead{Phase} & \colhead{Orbital Velocity (km s$^{-1}$)} & \colhead{S/N}
}
\startdata
\hline
\colhead{STELLA 2011 October--November}\\
\hline
55858.17166 & 0.033 & 77.268 & 188 \\
55859.16930 & 0.084 & 73.286 & 148 \\
55860.12126 & 0.133 & 66.652 & 156 \\
55861.12149 & 0.184 & 57.383 & 179 \\
55862.20816 & 0.239 & 45.733 & 162 \\ 
55863.20160 & 0.290 & 34.803 & 173 \\
55864.10758 & 0.336 & 25.562 & 179 \\
55866.20674 & 0.443 & 10.965 & 167 \\
55867.10558 & 0.489 &  8.871 & 130 \\
55868.21258 & 0.546 & 10.202 & 126 \\
55870.09460 & 0.642 & 21.615 & 156 \\
55871.09325 & 0.693 & 31.185 & 127 \\
55872.19978 & 0.749 & 43.194 & 134 \\
55873.18790 & 0.799 & 53.986 & 107 \\
55874.09618 & 0.846 & 63.003 & 201 \\
\hline
\colhead{UVES 2011 November}\\
\hline
55870.35375 & 0.655 & 23.922 & 457 \\ 
55872.33174 & 0.756 & 44.658 & 413 \\
55873.33838 & 0.807 & 55.563 & 496 \\
55885.34466 & 0.420 & 13.117 & 355 \\
55888.31713 & 0.571 & 12.198 & 412 \\
55888.31956 & 0.571 & 12.210 & 397 \\
55889.26356 & 0.619 & 18.098 & 382 \\
55889.26521 & 0.620 & 18.110 & 482 \\
55889.26614 & 0.620 & 18.118 & 576 \\
55893.30238 & 0.826 & 59.228 & 491 \\
\hline
\colhead{UVES 2012 November--December}\\
\hline
56239.34528 & 0.478 &  9.109 & 489 \\
56241.34542 & 0.580 & 13.117 & 486 \\
56242.30050 & 0.629 & 19.575 & 348 \\
56244.32863 & 0.733 & 39.633 & 523 \\
56246.32737 & 0.835 & 60.950 & 611 \\
56249.32214 & 0.987 & 77.918 & 458 \\
56254.29603 & 0.241 & 45.349 & 428 \\
56275.25630 & 0.310 & 30.597 & 423
\enddata
\label{spectra_log}
\end{deluxetable*}

\subsection{Photometry}

Differential $B$ and $V$ light curves of $\sigma$~Gem with comparison star HD 60318 and check star $
\upsilon$~Gem (HD 60522) were obtained by the Tennessee State University T3 0.4 m Automated Photometric Telescope (APT; Fairborn Observatory, USA) and are presented in Table 4 of \citet{roe15}.  Here, we use the photometry spanning 2011 October 5$-$2012 January 31 and 2012 October 10$-$November 30.  

\begin{figure}
 \hspace{-0.75cm}
  \includegraphics[angle=0,scale=.35]{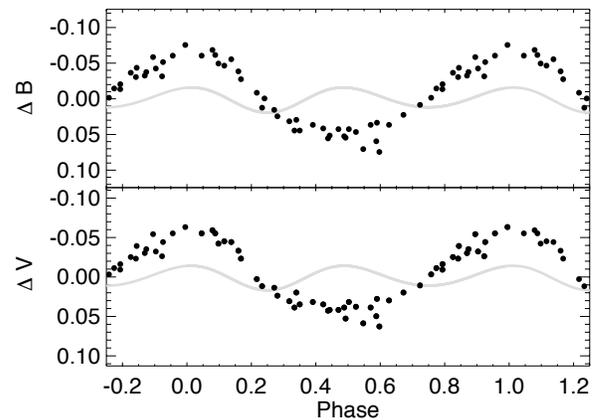} 
 \vspace{-1cm}
  \caption{Folded 2011 APT differential light curve of $\sigma$~Gem for $B$ and $V$ magnitudes.  The data points are the black circles.  The light grey curve in the background is the ELC-derived model for the $\sigma$~Gem light curve with no starspots that assumes a gravity darkening coefficient of $\beta = 0.02$, which was the best-fit coefficient determined in \citet{roe15}.}
  \label{2011LC}
\end{figure}

\begin{figure} 
 \hspace{-0.75cm}
  \includegraphics[angle=0,scale=.35]{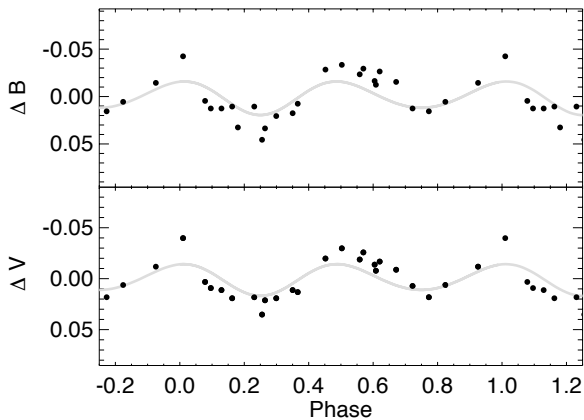} 
   \vspace{-1cm}
  \caption{Folded 2012 APT light curve of $\sigma$~Gem for differential $B$ and $V$ magnitudes as in Figure \ref{2011LC}.  }
  \label{2012LC}
\end{figure}

Due to insufficient phase coverage, we require more than one rotation period for the light curve.  The spot structures are assumed to remain stable over the length of the light curve.  These excerpt light curves are plotted in Figures \ref{2011LC} and \ref{2012LC}, phase-folded over the rotational period, $P_\mathrm{rot} = 19.6027$ days.  

In the 2012 data set, at phase $\phi = 0.010$ there is an outlier in both the $B$ and $V$ bands.  This data point was obtained on 2012 October 10, but we do not have any overlapping interferometric or spectroscopic data to investigate if this data point was part of a flare.  We do not exclude the point.  


\section{Imaging Methods}

Many studies have been published of spotted stars imaged with light curve inversion or Doppler imaging techniques individually \citep[e.g.,][]{sav08,kov15}, or comparing contemporaneous data sets  \citep[e.g.,][]{roe11}.  To date, two works have shown interferometrically imaged spotted stellar surfaces \citep{roe16,par15}.  No study has compared the simultaneous images of interferometric, spectroscopic, and photometric data sets on spotted stars.  We present the first such comparison here.

\subsection{Aperture Synthesis}

Aperture synthesis imaging is used on interferometric data with sufficient $uv$ coverage (the projection of the baselines onto the plane of the sky). 
Here, we use the imaging algorithm SURFace imagING (SURFING), which was created for the purpose of imaging interferometric data directly onto the surface of a rotating  spheroid \citep[Monnier in preparation;][]{roe16}.  SURFING treats the entire interferometric dataset as an ensemble of measurements in order to create a surface map.  Ideally, each pixel on the surface is the result of several overlapping observations.  SURFING allows for the elimination of degeneracies experienced by other imaging methods, providing independent measurements of some stellar parameters.  In a process similar to that  used in \citet{roe16} but accounting for the binary component, we ran SURFING using the input parameters found in Table \ref{params}.  During image reconstruction, we used a prior for the pixel values of a downward exponential with a maximum pixel value of $100\%$ and a decrease in surface brightness by a factor of $1/e$ for every $10\%$ in lower surface brightness (see Figure \ref{sigGemrot} in Appendix B).

\begin{deluxetable*}{l c}
\tabletypesize{\scriptsize}
\tablecaption{Input Parameters for Imaging of $\sigma$~Gem}
\tablewidth{0pt}
\tablehead{
\colhead{Interferometric Aperture Synthesis Imaging Parameters} & \colhead{Value} 
}
\startdata
primary major-to-minor axis ratio & $1.022$\\
inclination, $i$ ($^\circ$)$^a$ & $107.37$\\
ascending node, $\Omega$ ($^\circ$) & $1.1$\\
period, $P_\mathrm{orb}$ (days) & $19.6030$\\ 
time of nodal passage, $T_0$ (MJD)$^b$ & $53583.480$\\ 
limb-darkened disk diameter, $\theta_{\mathrm{LD}}$ (mas) & $2.425$\\ 
limb-darkening coefficient (power law) & $0.275$\\
$H$-band flux ratio & 252\\
separation, $a$ (mas) & 4.68\\

\hline
\colhead{Doppler Imaging Parameters} & \colhead{} \\
\hline
effective temperature, $T_\mathrm{eff}$ (K) & $4530$\\
inclination, $i$ ($^\circ$)$^a$ & $72$\\
metallicity, Fe/H & $0.0$\\
surface gravity, $\log g$ & 2.5\\
rotational velocity, $v \sin i$ (km s$^{-1}$) & 24.8 \\
orbital period, $P_\mathrm{orb}$ (days) & 19.6027\\
time of nodal passage, $T_0$ (MJD)$^{b,c}$ & 53583.61\\
microturbulence (km s$^{-1}$) & 0.8 \\
macroturbulence (km s$^{-1}$) & 2.0 \\

\hline
\colhead{Light-curve Inversion Imaging Parameters} & \colhead{} \\
\hline
photospheric temperature, $T_\mathrm{phot}$ (K)$^d$ & 4530\\
spot temperature, $T_\mathrm{spot}$ (K)$^e$ & 3800\\
inclination, $i$ ($^\circ$)$^a$ & 72.3 \\
limb-darkening coefficients $V$-band$^f$ & 0.767, 0.059\\
limb-darkening coefficients $B$-band$^g$ & 0.851, 0.158
\enddata
\tablecomments{$^a$Unlike interferometric imaging, Doppler and light curve inversion imaging cannot distinguish the star's orientation on the sky.  An inclination equivalent to that observed for $\sigma$ Gem and below $90^\circ$ are used for these methods.   \\
$^b$Time of maximum recessional velocity of the primary star.  \\
$^c$See Appendix C.\\
$^d$\citet{roe15}\\
$^e$\citet{kov15}\\
$^f$For $V$-band LI uses a square-root limb-darkening law and limb-darkening coefficients from \citet{van93}.\\
$^g$For $B$-band LI uses a logarithmic limb-darkening law and limb-darkening coefficients from \citet{van93}.}
\label{params}
\end{deluxetable*}

 \begin{figure*}
   \hspace{0cm}
  \includegraphics[angle=0,scale=0.75]{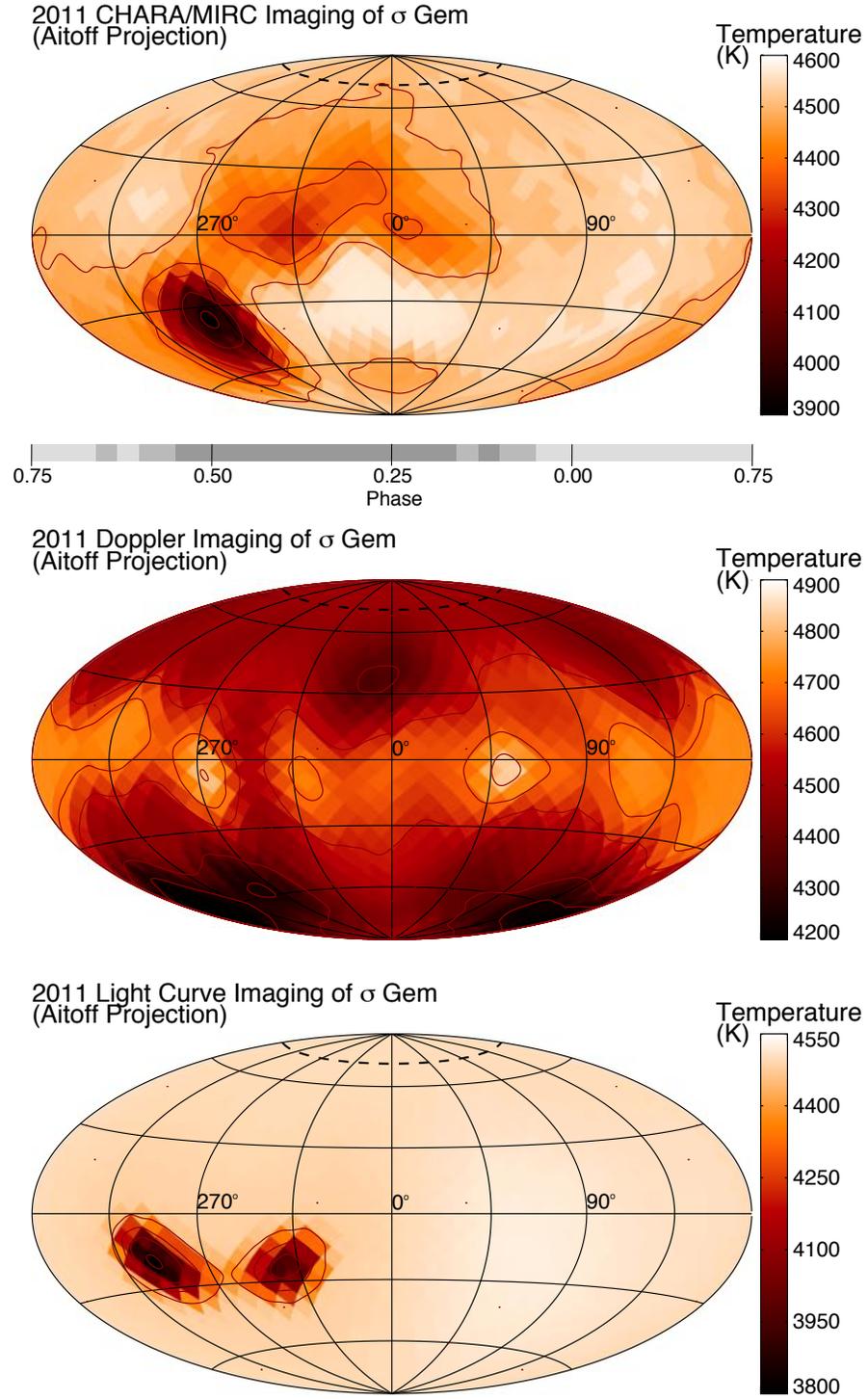} 
    \caption{Temperature maps of $\sigma$~Gem in the 2011 observing season.  The Aitoff projections show the surface temperature with the appropriate color scale to the right of the projection.  The latitudes hidden by stellar inclination appear above the dashed line around the top pole.  Top:  For the aperture synthesis interferometric map, each contour represents 100 K.  The gray bar beneath the Aitoff projection represents the number of times each phase was observed with MIRC with darker grays indicating more observations.  For 2011, each phase was observed 1--3 times.   Phases are noted on the gray bar.  For reference, at phase = 0.00, $90^\circ$ is at the center of the stellar disk; as time advances to phase = 0.25, $0^\circ$ is now at the center.
  Middle:  For the Doppler map, each contour represents 100 K.   Bottom:  For the light-curve inversion map, each contour represents 150 K.       }
    \label{Aitoff2011}
\end{figure*}

 \begin{figure*}
   \hspace{0cm}
  \includegraphics[angle=0,scale=0.75]{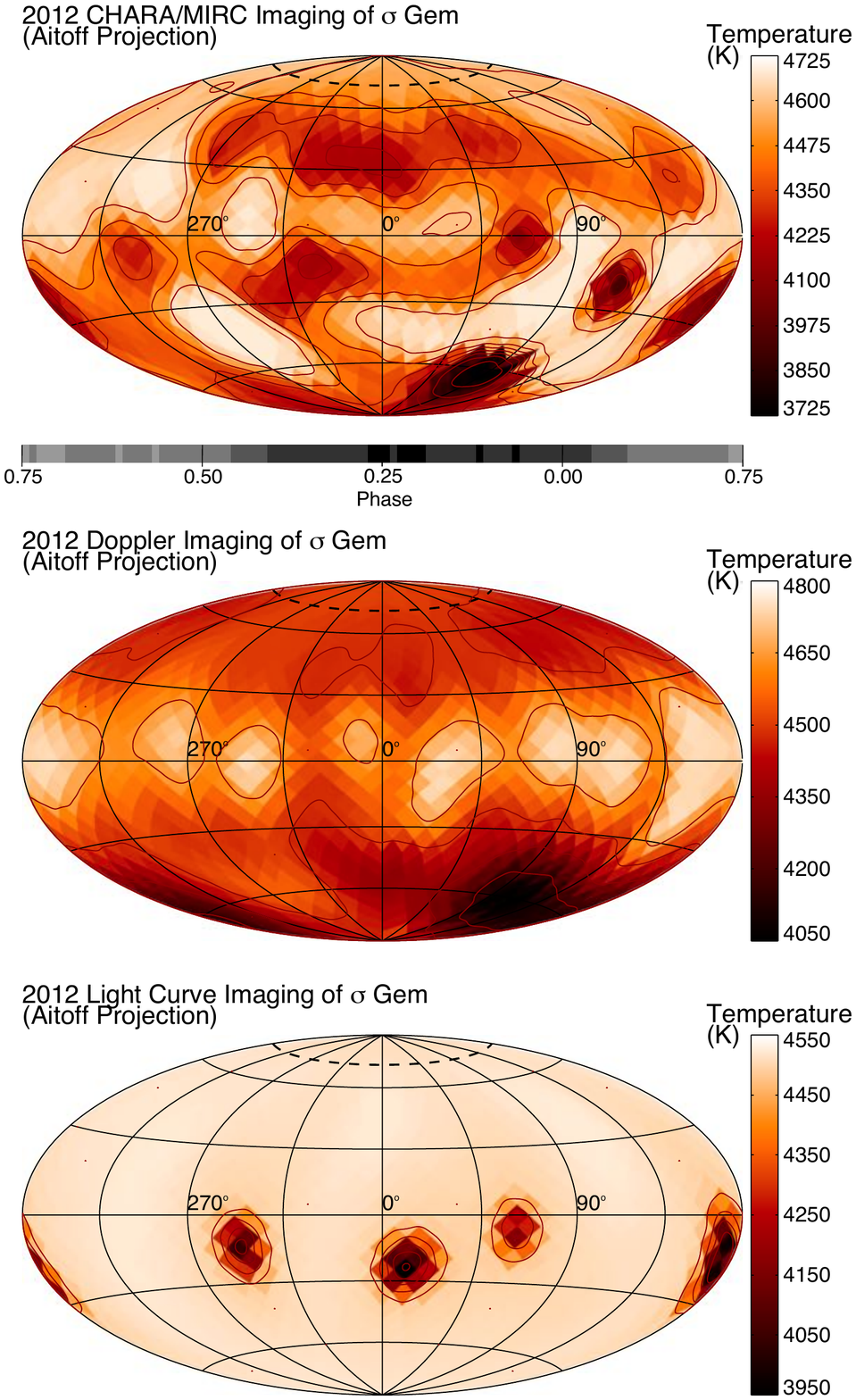} 
    \caption{Temperature maps of $\sigma$~Gem in the 2012 observing season as plotted in Figure \ref{Aitoff2011}.  Note that the contours are every 125 K for the interferometric map, every 150 K for the  Doppler map, and every 100 K for the light-curve inversion map in this figure. The gray bar beneath the interferometric map is similar to that used in Figure \ref{Aitoff2011}, but here each phase was observed 3--7 times.}  
  \vspace{1cm}
    \label{Aitoff2012}
\end{figure*}

$\sigma$~Gem is a partially Roche-potential-filling star with a ratio of equatorial to polar radius observed to be $1.02\pm0.03$ \citep{roe15} and modeled with the light curve modeling code Eclipsing Light Curve \citep[ELC;][]{oro00} to be 1.03.  Using the observed value in SURFING, we estimate the shape of $\sigma$~Gem as a prolate spheroid.  We use 768 tiles of equal surface area covering  $\sigma$~Gem, equivalent to 0.024 mas$^2$ in spatial resolution.  Each tile's value is a combination of all of the nights on which that region of the star was observed.  The results of our aperture synthesis imaging with SURFING are presented as temperature maps in Figures \ref{Aitoff2011} and \ref{Aitoff2012} and in $H$-band images in Figure \ref{sigGemrot} in Appendix B.  All phases of $\sigma$~Gem were observed at least once for these images.  In 2011, each phase has contributions from 1-3 nights of data, while in 2012, each phase has observations on 3-7 nights. We used the ten nights of observation in 2012 to test the reliability of the SURFING code.  The results and discussion are in Appendix B.   

For the interferometric temperature maps in Figures \ref{Aitoff2011} and \ref{Aitoff2012}, we convert from $H$-band intensities to temperatures assuming a Kurucz model \citep{cas04} and the average $H$ = 1.67.  With this, $\sigma$~Gem is observed to have temperatures listed in Table \ref{temps}.

\begin{deluxetable*}{l c c c c c c}
\tabletypesize{\scriptsize}
\tablecaption{Temperatures and Magnitudes of $\sigma$~Gem from Imaging}
\tablewidth{0pt}
\tablehead{
\colhead{Year} & \colhead{Image Type} & \colhead{$T_\mathrm{min}$ (K)} & \colhead{$T_\mathrm{max}$ (K)} & \colhead{$T_\mathrm{mean}$ (K)} & \colhead{$(T^4)_\mathrm{mean}^{1/4}$ (K)} & \colhead{$V_\mathrm{mean}$$^a$}
}
\startdata
2011 & light-curve inversion & 3880 & 4530 & 4490  & 4490 & 4.31\\
& Doppler & 4250 & 4840 & 4570 & 4570 & 4.20\\
& interferometric & 3950 & 4570 & 4490 & 4490 & 4.31\\
2012 & light-curve inversion & 3980 & 4530 & 4500 & 4500 & 4.30\\
 & Doppler & 4090 & 4750 & 4540 & 4550 & 4.24\\
  & interferometric & 3760 & 4690 & 4480 & 4490 & 4.32
\enddata
\tablecomments{$^a$$V = 4.14$ \citep{hen95}; $V = 4.29$ \citep{duc02}.}
\label{temps}

\end{deluxetable*}

The 2011 interferometric image of $\sigma$~Gem presents two strong starspot features near longitude $270^\circ$. 
We see one spot above and one below the stellar equator.  The 2012 image shows a more complex stellar surface with starspots peppering the surface.  We point out the dark features near the southern (visible) pole and the series of spots near the stellar equator.  We cannot determine if the spots present in 2011 have evolved into any of the features observed in 2012.

\subsection{Doppler Imaging}

We used the INVERS7PD inversion code developed by \citet{pis90}, and modified by \citet{hac01}. The code uses Tikhonov regularization and compares observations to a grid of local line profiles calculated with the SLOC5 spectral synthesis code \citep{ber91} and Kurucz model atmospheres \citep{kur93}. Atomic line parameters are obtained from VALD \citep{pis95, kup99}, while molecular line parameters are calculated as described by \citet{ber98}. The local line profiles were calculated for twenty limb angles, nine temperatures between 3500~K and 5500~K with a 250~K step, and wavelengths between 6408.5 -- 6441~{\AA} with a wavelength step of 0.01~{\AA}.  We divide the surface into a grid of 40 bands of latitude, each split into 80 longitude sections. 
Before inversion, the local line profiles are convolved with a Gaussian instrumental profile and a radial-tangential macroturbulence velocity. For the separate photometric output we use the same code and models but with a sparser wavelength grid ranging from 3600~{\AA} to 7350~{\AA} and step size of 50~{\AA}. The stellar parameters are fixed to the values given in Table \ref{params}.

In the inversions, iron and calcium lines in the wavelength region of 6410-6440~{\AA} are used (Fe I 6411~{\AA}, Fe I 6419~{\AA}, Fe I 6421~{\AA}, Fe I 6430~{\AA}, and Ca I 6439~{\AA}). These lines are  traditionally used for Doppler imaging. 
The inversions are done using all the five lines simultaneously. 

INVERS7PD does not account for the ellipsoidal shape of $\sigma$~Gem, and we do not remove the signature from the spectra, as the ellipsoidal variation signature is not significant enough to change the Doppler imaging results.  

For the 2011 inversions the STELLA data were used simultaneously with the UVES data. The UVES data do not have good enough phase coverage to be used alone, but they have better S/N and higher resolution than the STELLA data, and therefore they 
improve the stability of the resultant map. Before inversion, the UVES data were rebinned to the same resolution as the STELLA data.

The UVES data for 2012 were obtained over almost two stellar rotations. Most of the data, seven phases, are obtained within one rotation. There is only one data point that has been obtained one rotation later.  For testing whether spot evolution occurred within the extra rotation, inversions were carried out both with and without the spectrum obtained one rotation apart. No significant difference could be seen in the resultant map, nor in how well the model fit the observations. Therefore, all eight spectra obtained in 2012 were used in the Doppler images presented in Figures \ref{Aitoff2011} and \ref{Aitoff2012}.

The 2011 Doppler image shows a significant starspot region around the southern (visible) pole.  The feature is neither polar nor symmetric about the pole.  The equator is the brightest region of the star.  Another starspot is located in the northern hemisphere around longitude $0^\circ$.  The 2012 surface again shows the strong starspot near the southern pole with the brightest stellar material found around the equator.  Temperature contours indicate cooler regions in the northern hemisphere.  
The temperatures of this surface are included in Table \ref{temps}.

\subsection{Light-curve Inversion}

For inverting the $B$ and $V$ APT light curves, we use the Light-curve Inversion \citep[LI;][]{har00} algorithm.  LI is a non-linear inversion algorithm that breaks the stellar surface into patches that are approximately equal in surface area (60 bands of latitude with the equatorial bands broken into 90 patches), varying each independently as described in \citet{har00} and \citet{roe11}. 

LI used a spherical surface for $\sigma$ Gem, and we accounted for the deviation from a spherical star by removing the effect of ellipsoidal variations in the light curves.  To do so, we removed the \citet{roe15} ellipsoidal variation model light curve \citep[generated with ELC;][]{oro00} with the best-fit gravity darkening parameter ($\beta = 0.02$; see Figures \ref{2011LC} and \ref{2012LC}).  These adjusted light curves are the input for LI.     

The stellar parameters of $\sigma$~Gem assumed for LI are in Table \ref{params}.  The output of LI is relative intensities, which we linearly map from intensities to temperatures ranging from 3800 K to 4350 K, the assumed spot and photospheric temperatures, respectively.  The results are shown in Figures \ref{Aitoff2011} and \ref{Aitoff2012} with the temperatures included in Table \ref{temps}.

The 2011 LI image shows a simple stellar surface with two spots near longitude 270$^\circ$ and at approximately the same latitude.  The 2012 LI image shows more structure with four equatorial starspots.  Using LI with only two light curves, as we did here will limit the latitude information available in the inversions.  Simulations have shown that using light curves from four bandpasses will improve spot latitude information but will not constrain the values as well as the other imaging methods discussed above \citep{har00}.


\section{Comparison of Imaging Results}
 
\begin{figure*}
 \vspace{-0.5cm}
  \includegraphics[angle=0,scale=0.7]{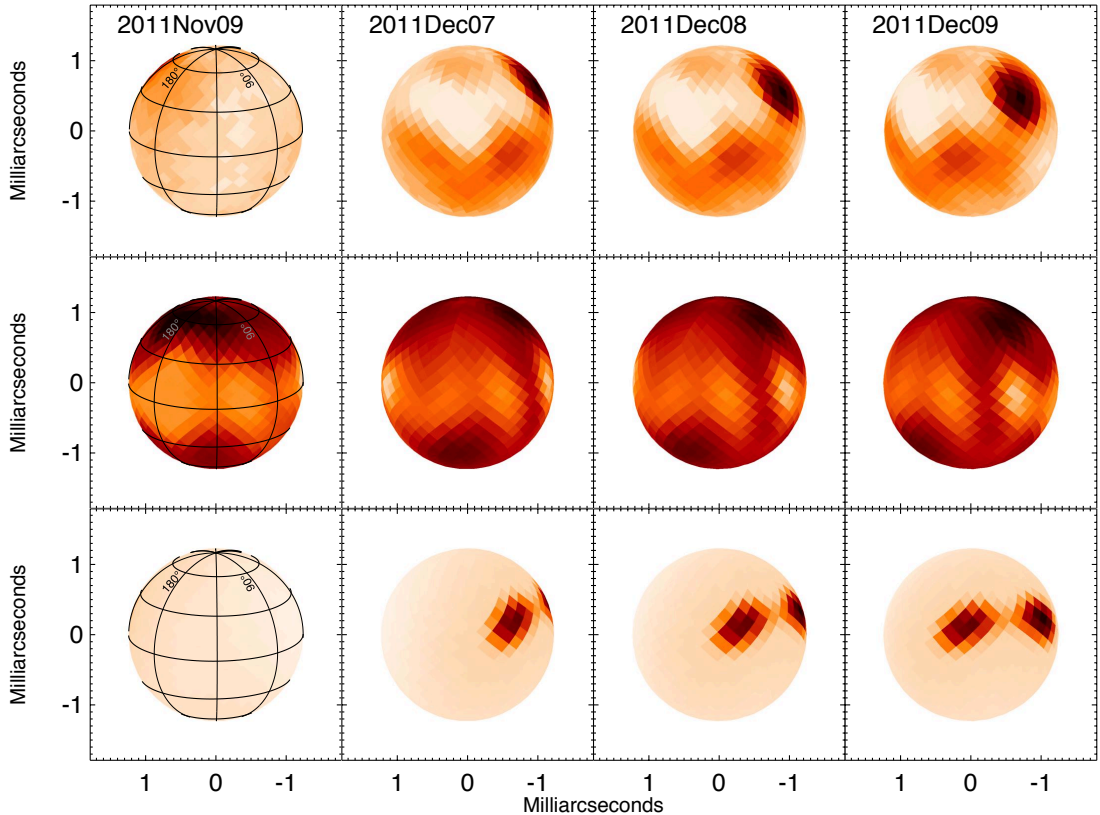} 
   \vspace{-2cm}
  \caption{Temperature maps of $\sigma$~Gem on the dates of the 2011 CHARA/MIRC observations for the interferometric aperture synthesis, Doppler, and light-curve inversion imaging methods.  The projections use the temperature maps appearing in Figure \ref{Aitoff2011}.  These surfaces are not plotted as they appear on the sky:  the visible (southern) rotational pole is plotted pointing upward and the star rotates from \textbf{right to left}.  Each column represents a night of CHARA/MIRC observations and is listed in the top row of images, which are those from SURFING.  The middle row are the contemporaneous Doppler maps, and the bottom row are the light-curve inversion maps.  }     
    \label{imagecompare2011}
\end{figure*}

\begin{figure*}
 \vspace{-0.5cm}
  \includegraphics[angle=0,scale=0.7]{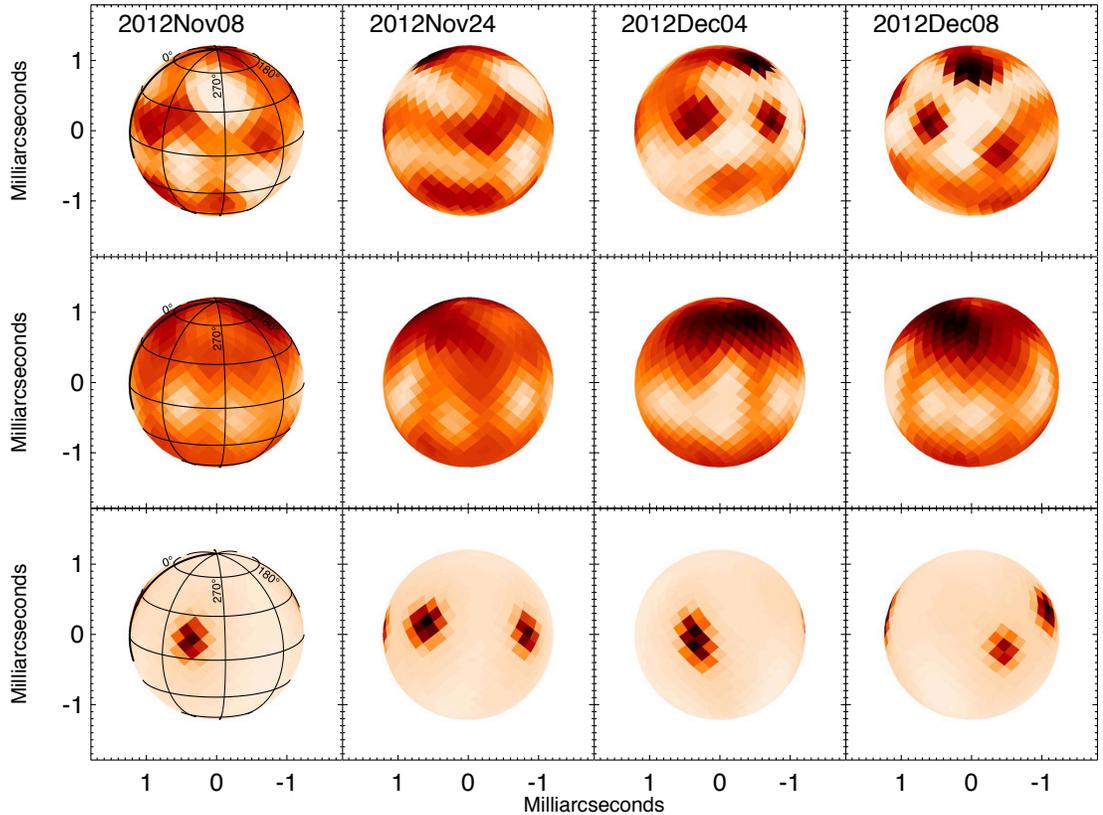} 
     \vspace{-2cm}
  \caption{Temperature maps of $\sigma$~Gem on the dates of the 2012 CHARA/MIRC observations for the interferometric aperture synthesis, Doppler, and light-curve inversion imaging methods.  The maps are as described in Figure \ref{imagecompare2011} using the temperature maps of Figure \ref{Aitoff2012}.}
    \label{imagecompare2012}
\end{figure*}

 In order to do a direct comparison of the stellar surface as observed with the three different imaging techniques, we present $\sigma$~Gem as it appeared on the dates of the interferometric observations (see Figures \ref{imagecompare2011} and \ref{imagecompare2012}).   We show all four nights of observation in the 2011 observing season and four of the ten nights observed in 2012.  We draw attention here to specific surface features and their presence or absence using the different imaging techniques.    
 
\begin{figure}
\hspace{-1.5cm}
\includegraphics[angle=90,scale=.4]{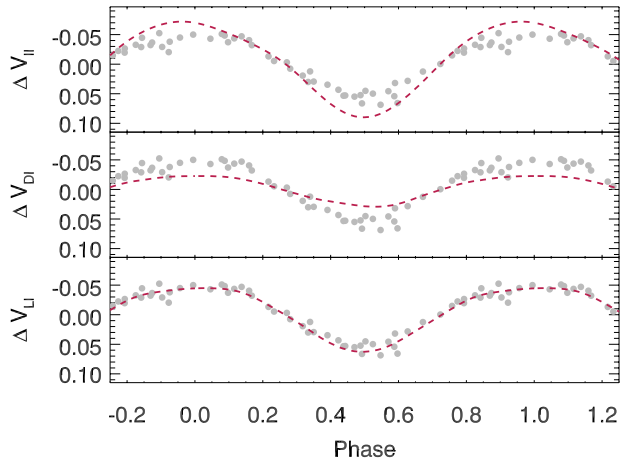} 
\vspace{-1cm}
  \caption{Comparison of observed and reconstructed light curves.   The folded 2011 APT $V$-band light curve of $\sigma$~Gem with the signature of the ellipsoidal variations having been removed appears as gray circles in each panel.  The red dashed line plotted in each panel is the reconstructed light curve from the surfaces presented in Figure \ref{Aitoff2011}, which are created from the interferometric data directly (top) and the Doppler (middle) and light-curve inversion (bottom) imaging results.    }
  \label{allLC2011}
\end{figure}

\begin{figure}
 \hspace{-1.5cm}
  \includegraphics[angle=90,scale=.4]{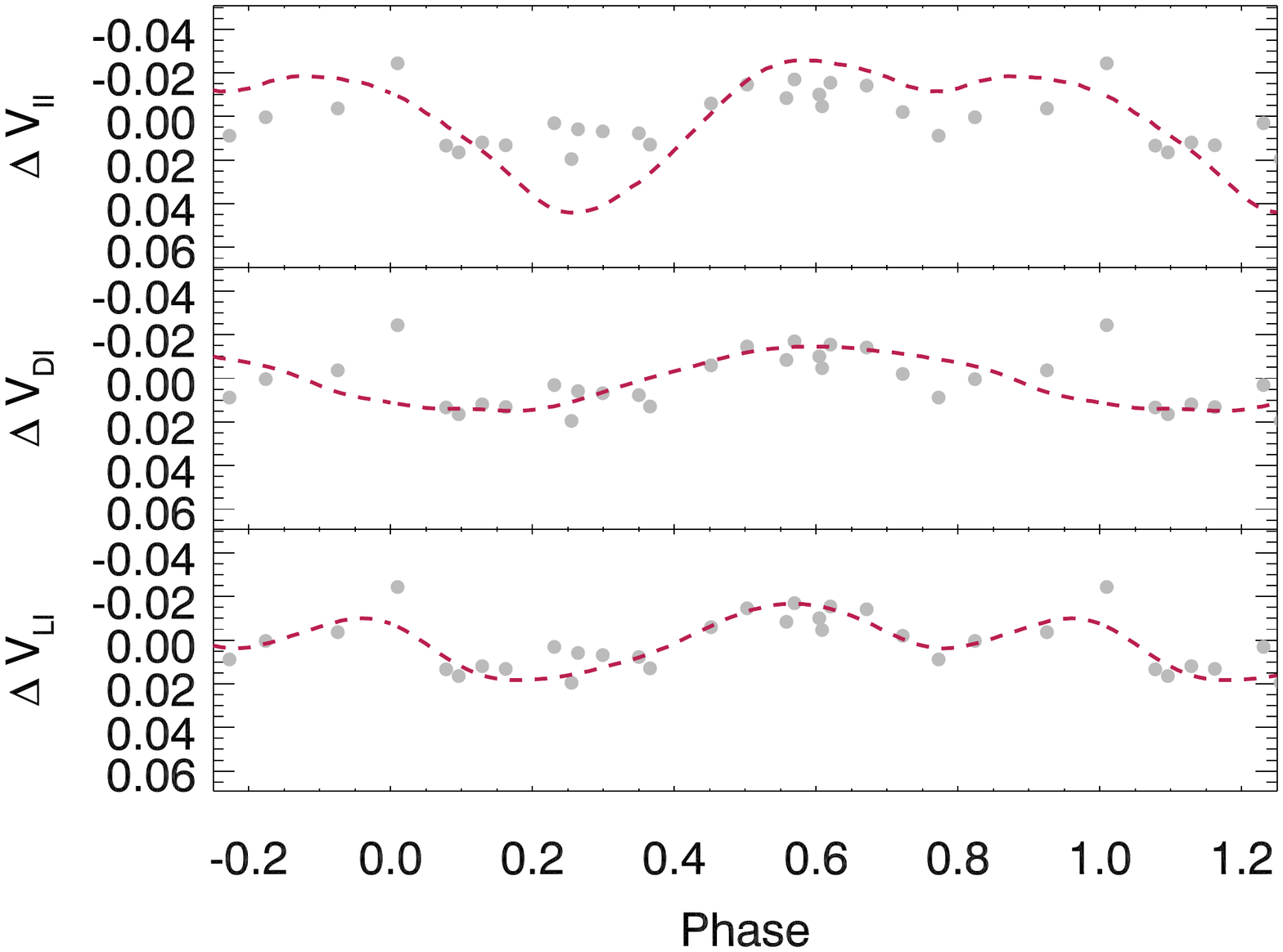}  
  \vspace{-1cm}
  \caption{Comparison of observed and reconstructed light curves.  These plots of the 2012 data and results are as in Figure \ref{allLC2011}.   }
  \label{allLC2012}
\end{figure}

The temperature maps of Figures \ref{imagecompare2011} and \ref{imagecompare2012} were plotted on the ellipsoidal surface of $\sigma$ Gem.  The temperatures were then combined with Kurucz models \citep{cas04} to determine flux in the $V$ bandpass and to create the light curves.  To obtain the light curves presented in Figures \ref{allLC2011} and \ref{allLC2012}, we removed the ellipsoidal variations, as we did with the observed light curves.
 
In both data sets, we note that the images from the different reconstructions are not completely consistent.  However, the reconstructed light curves match well with the observed light curves.  We note the amplitude of the Doppler imaging light curve is smaller than that observed, while the interferometric imaging light curve amplitude is larger.  The reconstructed light curves indicate that the methods are reliable in their determinations of the longitudes of the spots.  The biggest differences between the models comes from the latitude of the starspots, a value for which limited information is available to Doppler imaging and even less to light curve inversion imaging.

Table \ref{temps} gives the minimum, maximum, and mean temperatures of each temperature map.  
Interferometric imaging is capable of resolving the surface temperature profile as a function of latitude, so temperature can be measured at each latitude and longitude of the surface.  This is not the case in Doppler imaging where the mean temperature profile is not variable in latitude.  The temperature is dependent upon the absorption line profile modeling, which is dominated by the hottest regions of the star and skews the stellar temperature to higher values.  
Light-curve inversion provides even less information for the stellar temperature; the temperatures for our LI results were prescribed as input based on the stellar temperature used in \citet[][which is an average value based on spectroscopically-derived temperatures found in the literature]{roe15} and the spot temperature assigned by \citet[][determined with Doppler imaging]{kov15}.  

\subsection{2011 Images}

During the 2011 observations, $\sigma$~Gem exhibited a single strong feature in the light curve (see Figure \ref{allLC2011}).  This feature was imaged as two spots which are closely located in longitude (around $270^\circ$) in the interferometric and photometric images (see, in particular, the fourth column of Figure \ref{imagecompare2011}).  It is clear here that the light-curve inversion method is very limited in its ability to determine the latitudes of the spots when using only two bandpasses.  The Doppler image exhibits the two spots at similar longitudes to those seen in the other results, but they appear at more extreme latitudes than in the interferometric image; this could be attributed to the difficulty of imaging spots in the less visible hemisphere for Doppler imaging.  

Also in the Doppler image, the southern (visible) pole of $\sigma$~Gem is seen to have a high-latitude, nearly-circumpolar spot centered around a longitude of $180^\circ$ (see the first column, middle row of Figure \ref{imagecompare2011}).  This feature is not observed in the interferometric or light curve images.  While the interferometric observations are more sparse on this side of the star with these phases only observed once, the light curve gives no indication of spot features on this side of the star.  The spot is of high-latitude such that most of it would always be visible, and thus would be undetected by light-curve inversion.

\subsection{2012 Images}

The observed, somewhat featureless light curve  shows evidence of a stellar surface covered in several dark and bright spots that nearly ``cancel'' each other out as the star rotates (see Figure \ref{imagecompare2012}).  The latitudes of the spots pictured here show significant discrepancies, while the longitudes tend to agree. 

The interferometric and Doppler images show southern (visible) hemisphere features and reveal evidence of the high-latitude spot below $-30^\circ$.  However, the equatorial spot structures evident in the interferometric image are fainter in the Doppler image.  There is evidence of dark and bright regions along the Doppler image's equator, but they are not as cool as the other spots present.  

The photometry reproduces four of the equatorial starspots seen on the interferometric surface.  The discrepancy between the number of equatorial starspots of the interferometric aperture synthesis image and that of the light-curve inversion is likely due to the latter method not resolving all of the spots.  In the light-curve inversion image, the starspots at the more extreme interferometric latitudes are not reproduced. 

Light-curve inversion is unable to reconstruct features that do not rotationally modulate.  With so many features on the surface, the light-curve inversion can only be a simplification of the actual surface.  While Doppler imaging is capable of distinguishing spots at different latitudes because their effects on the line profiles are different, the ability to distinguish these features depends upon the quality of the data and errors in the line profile modeling.  The combination of these effects will cause a blurring of the features in Doppler imaging.  


\section{Conclusions}
 
In this paper, we investigated the first comparisons of three different imaging techniques.  
We present two unprecedented data sets of simultaneous interferometric, spectroscopic, and photometric observations.  With these datasets, we aimed to compare the results of three state-of-the-art imaging techniques in order to validate the methods against each other.  Despite deviations between the resultant images, the agreement is good between the synthetic light curves created from the images and the observed light curve.  This agreement emphasizes that the starspot longitudes determined by the different imaging techniques are reliable.  However, the large differences in the starspot latitudes emphasize the shortcomings of the imaging methods.  
 
The limitations of light-curve inversion are highlighted in our comparison images, as light-curve inversion is only able to reconstruct simple surface features and is unable to distinguish between hemisphere (see, for example, the latitude of the starspots in the 2011 images, as in Figure \ref{imagecompare2011}).  Light-curve inversion is additionally limited for a complicated spot structure, such as the 2012 surface.  The many bright and dark regions, as seen in the interferometric and Doppler images will contribute to the light curve such that they ``cancel'' each other out during rotations, muting the detected features.  The resultant surface will be much simpler than reality.  In particular, applying LI to light curves of two bandpasses does not allow for constraints to be placed on spot latitudes.  Doppler imaging, however, is better able to image complicated surfaces.  While the strong features are detected by Doppler imaging, many spots moving across the surface can be simplified here, too, if the data are not of high enough quality.  However, unlike our two-bandpass light-curve inversion, Doppler imaging is capable of obtaining spot latitude information, especially in the more visible hemisphere.

We note, in particular, that the direct imaging of stellar surfaces available with optical interferometry is unmatched in its ability to capture the surface as it appears on the sky.  The complicated surface features present on the interferometric image of $\sigma$~Gem in 2012 emphasize the importance of understanding that the results from Doppler and light-curve inversion imaging could be oversimplified.  On the other hand, a major limitation of aperture synthesis imaging is the restriction imposed by only having data from a small number of telescopes in fixed positions.  While SURFING applied to MIRC data is able to image individual surface features well, it may fail to reveal large features like the smooth temperature gradients.  These gradients may not produce large enough asymmetries on the stellar surface to be detected or would only be detected by baselines not available at the CHARA Array.  Whether such smooth variations across the surface are actual features or artifacts and what interferometric observations would be required calls for simulations that are outside the scope of this paper.  

Such temperature gradients are seen in Doppler imaging, but we are presently unable to verify them interferometrically or determine if they are artifacts of the limited capabilities of Doppler imaging to accurately reconstruct surface temperatures.  As mentioned in Section 4, Doppler imaging depends upon absorption line modeling that is skewed to the highest temperatures.  It is possible that these temperature gradients with dark regions near and at the poles are the result of errors in fitting the spectra with model atmospheres or insufficient phase coverage. 
Further studies are necessary to determine the origin of these features. 

The temperatures of the aperture synthesis and light-curve inversion images, as seen in Figures \ref{Aitoff2011} and \ref{Aitoff2012}, as well as Table \ref{temps}, are lower than those of the Doppler images.  The spectroscopic observations are less-sensitive to high-temperature features, which can introduce hot artifacts on the surface, resulting in artificially increased average temperatures \citep[e.g.,][]{som15}, resulting in the higher Doppler image temperatures presented in Table \ref{temps}.  While the temperatures and resultant magnitudes for all of the imaging methods are consistent with literature values, further efforts to separate the photospheric and spot temperatures to accurately determine the stellar parameters of spotted stars are necessary \citep[e.g.,][]{gul17}.
 
\section*{Acknowledgements}

The interferometric data in this paper were obtained at the CHARA Array, funded by the National Science Foundation through NSF grants AST-0908253 and AST-1211129, and by Georgia State University through the College of Arts and Sciences.  The MIRC instrument at the CHARA Array was funded by the University of Michigan.  
The Doppler imaging in this work was based on observations collected at the European Organisation for Astronomical Research in the Southern Hemisphere under ESO programme 288.D-5007 and 090.D-0312.  
The photometric data were supported by NASA, NSF, Tennessee State University, and the State of Tennessee through its Centers of Excellence program.
R.M.R.\ would like to acknowledge support from the NASA Harriet G.\ Jenkins Pre-Doctoral Fellowship and a Rackham Graduate Student Research Grant from the University of Michigan.  
J.D.M.\ and R.M.R.\ acknowledge support of NSF grant AST-1108963.  F.B.\ acknowledges funding from NSF awards AST-1445935 and AST-1616483.  

This research has made use of the CDS Astronomical Databases SIMBAD and VIZIER\footnote{Available at http://cdsweb.u-strasbg.fr/}, operated at CDS, Strasbourg, France and the Jean-Marie Mariotti Center \texttt{SearchCal}\footnote{Available at http://www.jmmc.fr/searchcal} and \texttt{Aspro}\footnote{Available at http://www.jmmc.fr/aspro} services
co-developed by FIZEAU and LAOG/IPAG.

\appendix

\section{A. Interferometric Observables}

\begin{figure}
 \hspace{3cm}
 \includegraphics[angle=0,scale=.45]{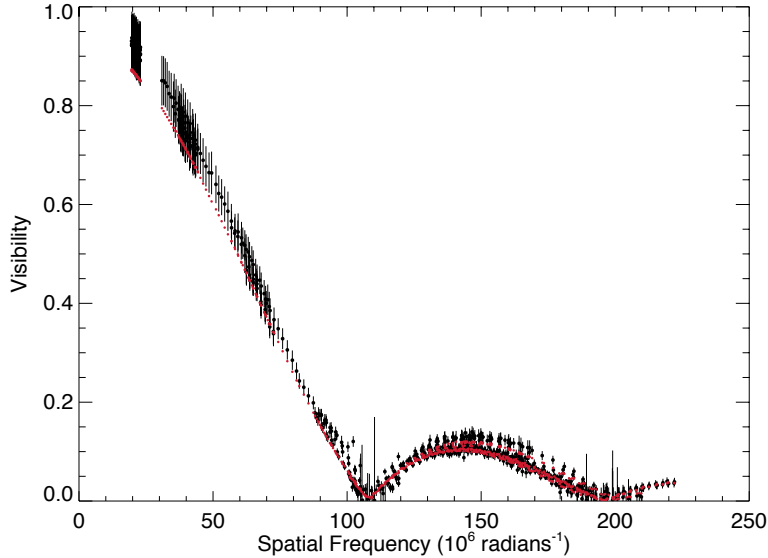} 
 \vspace{-1cm}
  \caption{Visibility curve of UT 2012 November 7 $\sigma$~Gem CHARA/MIRC observations.  The observed visibilities are plotted in black with 1$\sigma$ error bars.  The SURFING model visibilities (see Section 3.1) are overplotted in red.}
  \label{2012Vis}
\end{figure}

\begin{turnpage}
 \begin{figure*}
 \vspace{-2cm}
  \includegraphics[angle=90,scale=.88]{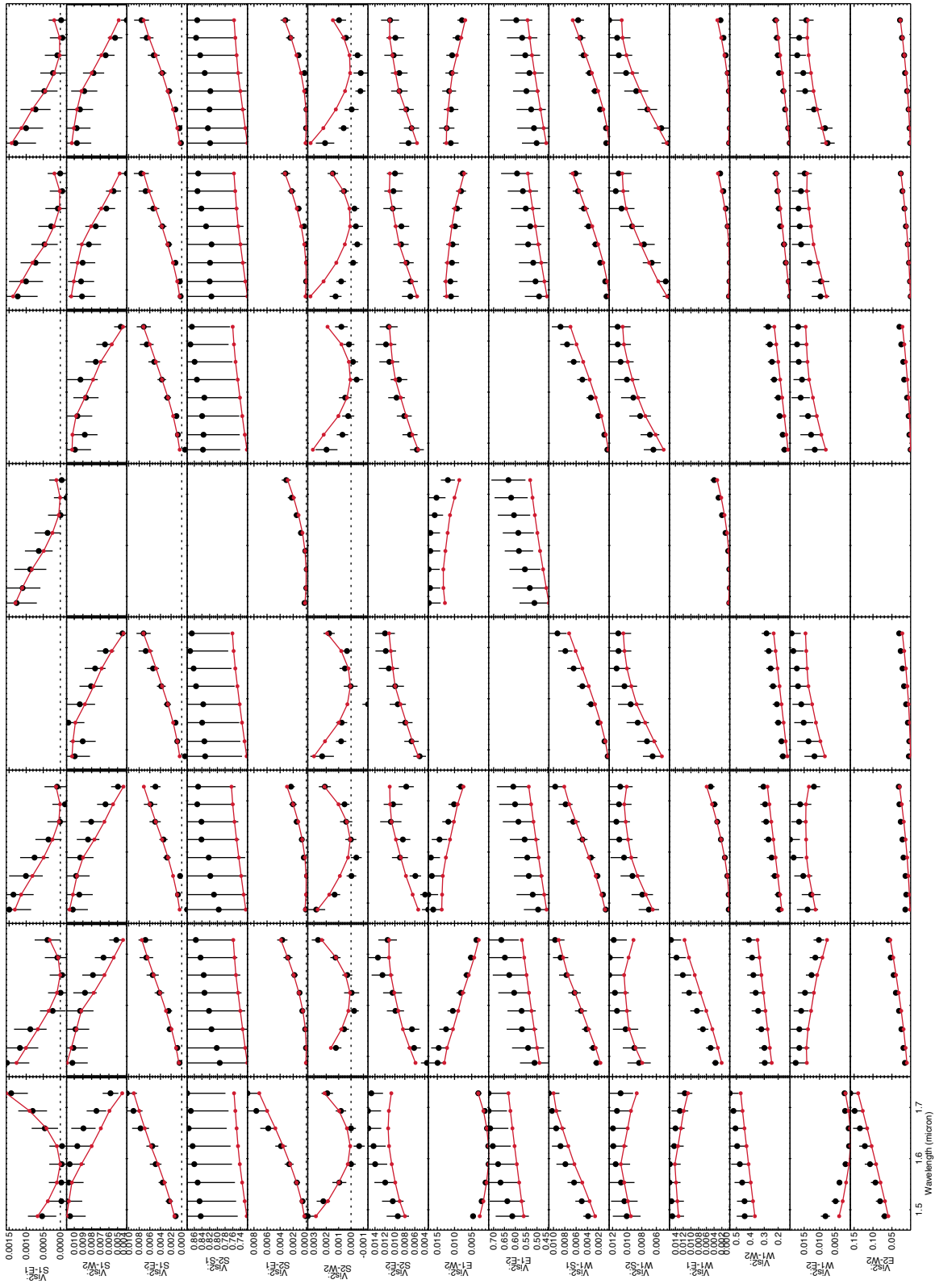}
  \caption{Squared visibilities of UT 2012 November 7 $\sigma$~Gem CHARA/MIRC observations.  Each block represents a pair of telescopes in a temporal block of observations.  The data are plotted with black circles, and the model is plotted in smaller red circles connected by a red line.}
  \label{2012V2}
\end{figure*}
\end{turnpage}

\begin{turnpage}
 \begin{figure*}
 \vspace{-2cm}
  \includegraphics[angle=90,scale=.88]{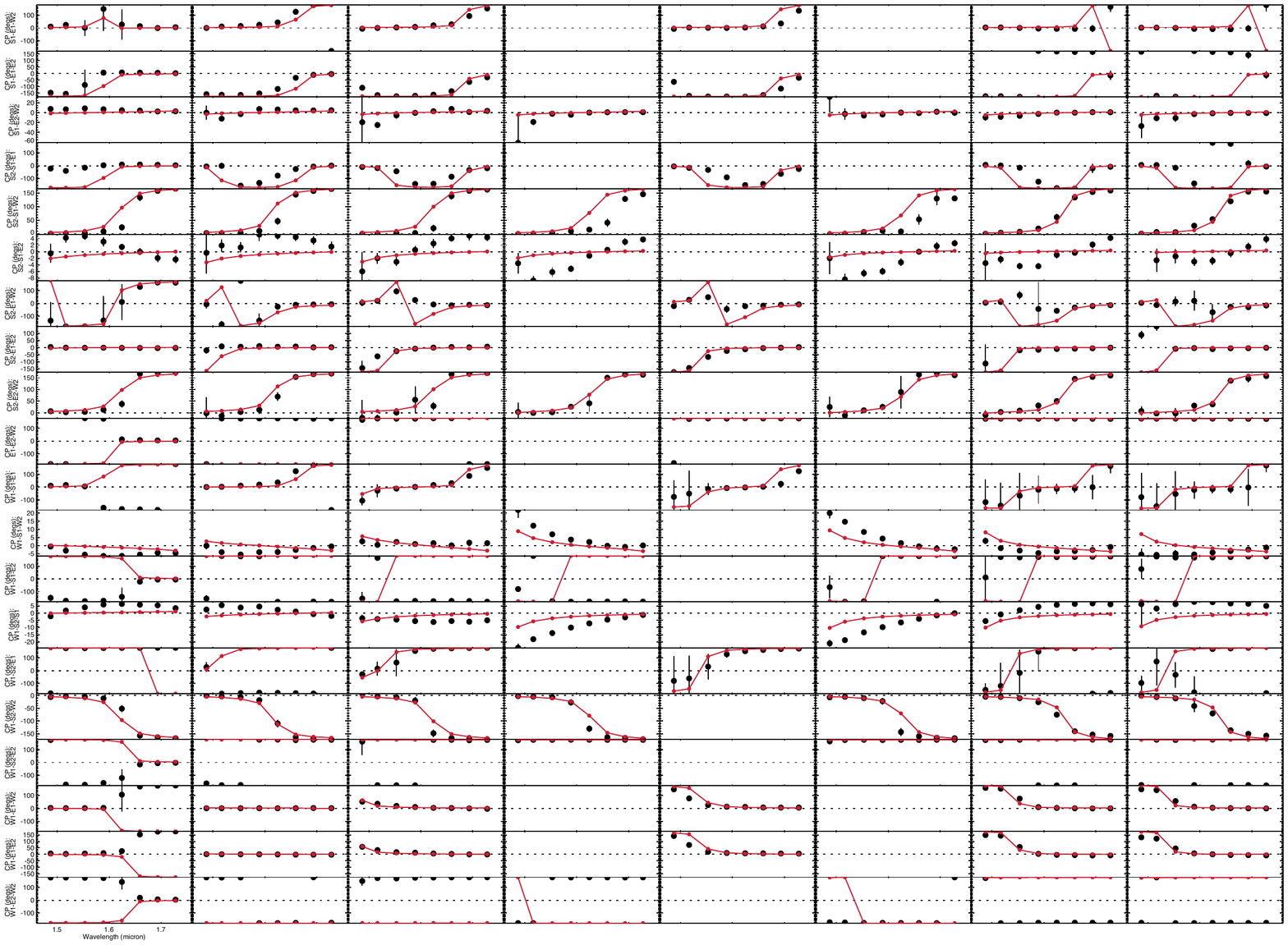} 
  \caption{Closure phases of UT 2012 November 7 $\sigma$~Gem CHARA/MIRC observations.  Each block represents a set of three telescopes in a temporal block of observations.  The data and model are as in Figure \ref{2012V2}.}
  \label{2012CP}
\end{figure*}
\end{turnpage}

\begin{turnpage}
\begin{figure*}
 \vspace{-2cm}
  \includegraphics[angle=90,scale=.88]{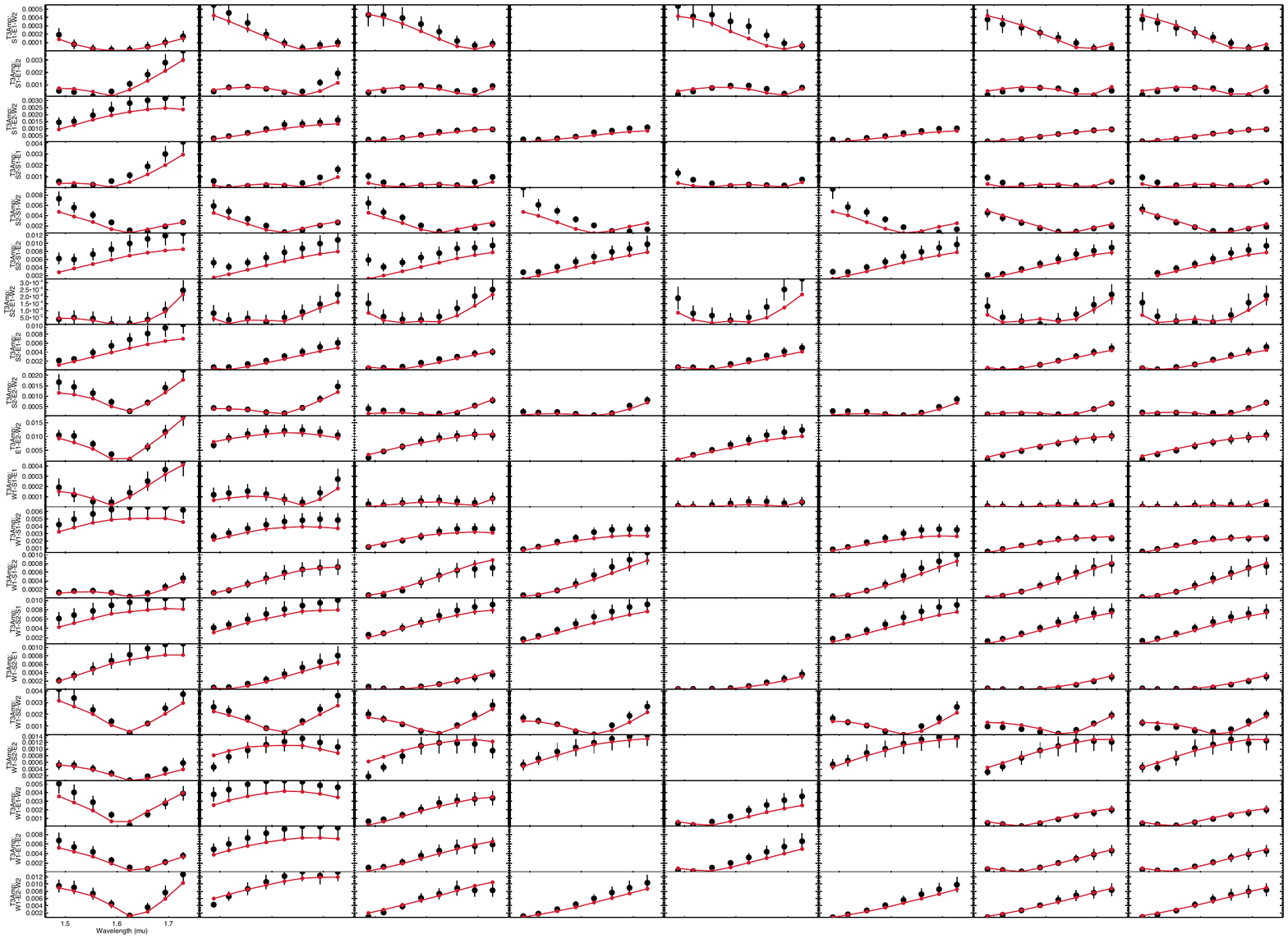} 
  \caption{Triple amplitudes of UT 2012 November 7 $\sigma$~Gem CHARA/MIRC observations.  Each block represents a set of three telescopes in a temporal block of observations.  The data and model are as in Figure \ref{2012V2}.}
  \label{2012T3}
\end{figure*}
\end{turnpage}

The interferometric observations acquired by MIRC at the CHARA Array included the observables of visibility, closure phase, and triple amplitude.  Examples of these observations are presented in Figures \ref{2012Vis} - \ref{2012T3}.  The data were reduced with the standard MIRC pipeline.

\section{B.  SURFING Results}

In Figures \ref{2012Vis} - \ref{2012T3}, we include the SURFING results for the example night.  The reduced $\chi^2$ of the squared visibilities is 1.75 and 1.72 for the 2011 and 2012 epochs, respectively.  For the bispectrum, the reduced $\chi^2$ values are 1.83 and 2.33 for 2011 and 2012, respectively.  

In order to balance imaging between fitting to noise and creating a smooth surface, we underestimate the error bars applied to the observables that are used in SURFING.  The effect of this underestimation results in increasing the reduced $\chi^2$ values, which is especially seen in fitting the closure phases.  The procedure used here is similar to that used for $\zeta$ Andromedae in \citet{roe16}, but here we account for the binary companion in SURFING. The $H$-band flux ratio between the primary and secondary stars of $\sigma$ Gem is smaller than that of $\zeta$ Andromedae.  For $\sigma$ Gem, the companion contributes $3\%$ of the $H$-band light of the system.

\begin{figure*}
 \hspace{0.5cm}
  \includegraphics[angle=0,scale=0.65]{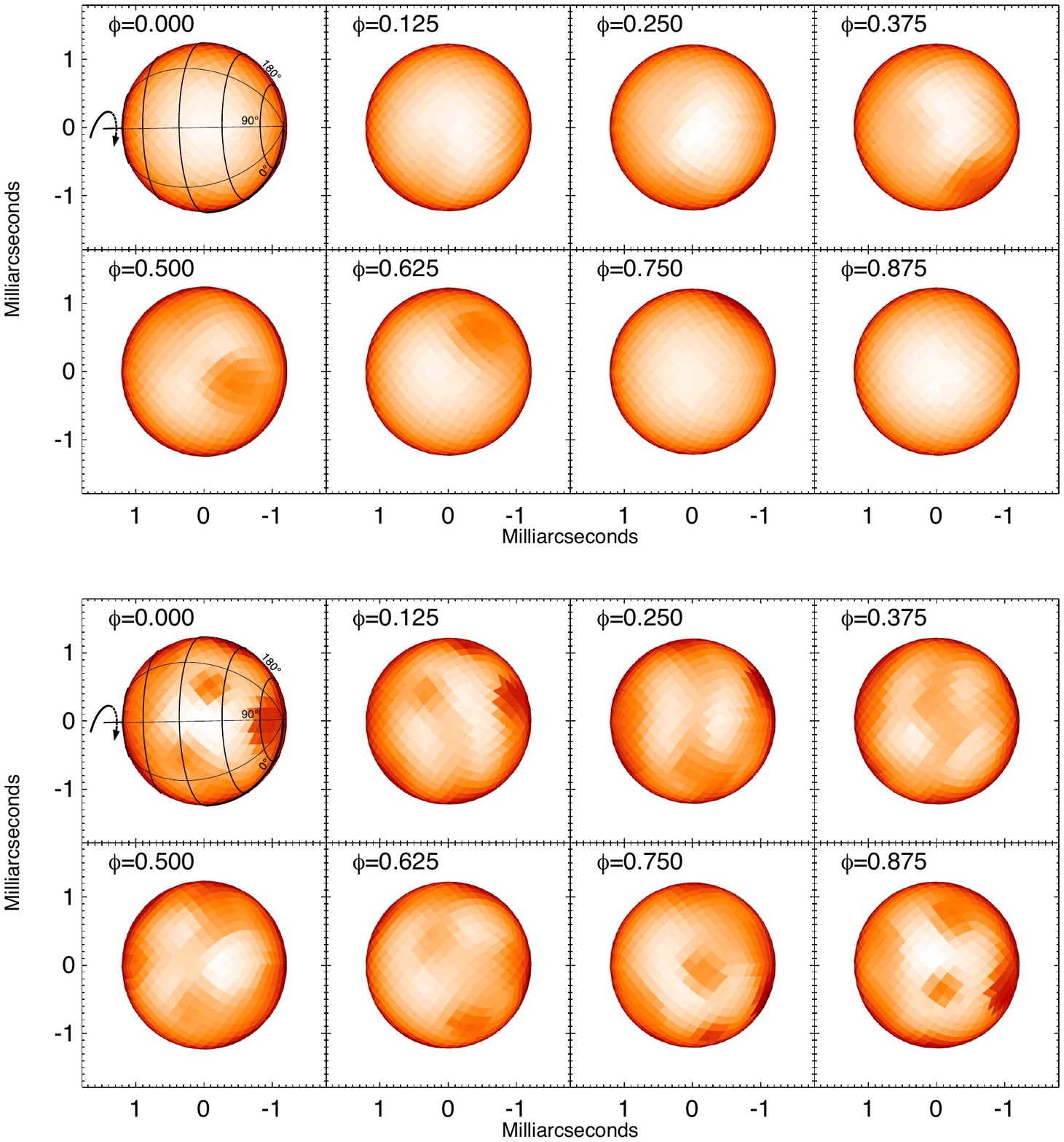} 
    \caption{Surface image of $\sigma$~Gem.  Top:  Eight views of $\sigma$~Gem as it appears on the sky in $H$-band in 2011 November - December. The designation of phase assumes radial velocity conventions for a circular orbit.  At phase $\phi = 0.000$, $0^\circ$ longitude is located at the bottom edge of the star with $90^\circ$ across the middle of this visible hemisphere.  As time advances, the longitude at the middle of $\sigma$~Gem decreases (at $\phi = 0.250$, the $0^\circ$ longitude is in the middle of the visible hemisphere of the star).  The images are oriented such that east is to the left and north is up.
  Bottom:  As above, but for 2012 November - December observations.}
  \label{sigGemrot}
\end{figure*}

We include images of $\sigma$~Gem as it appeared on the sky in $H$-band in both 2011 and 2012 (see Figure \ref{sigGemrot}).  These images include limb darkening.  

\begin{figure}
 \begin{center}
 \hspace{-1cm}
 \vspace{-2cm}
  \includegraphics[scale=.75]{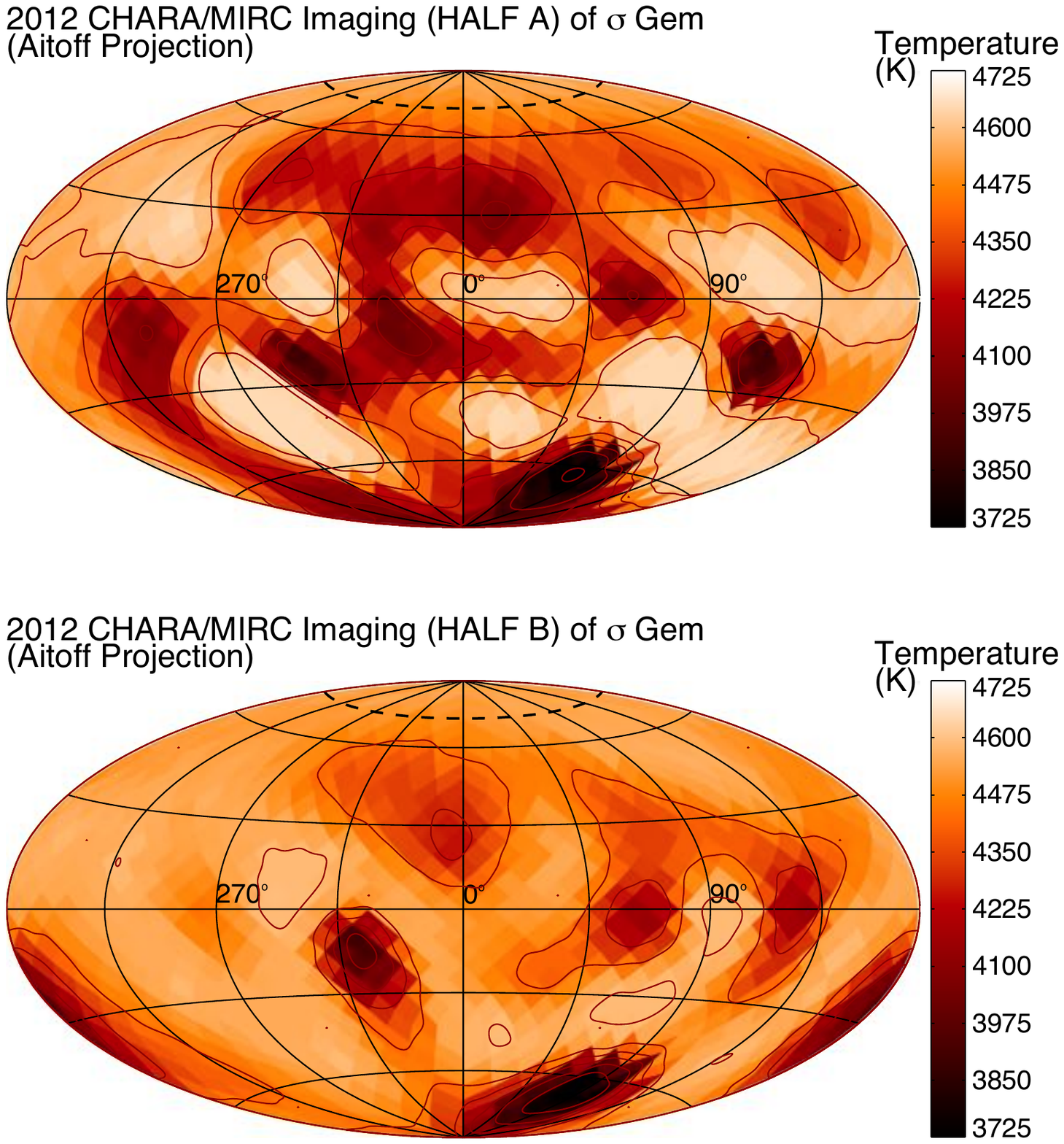} 
  \vspace{-2cm}
  \caption{Comparison of the 2012 interferometric temperature map divided into two data sets.  These Aitoff projections are as described in Figure \ref{Aitoff2012}, top.  Half A consists of nights 2012 November 7, 21, 24; December 4, and 7.  Half B consists of nights 2012 November 8, 22, 25; December 5, and 8.}
  \label{2012SURFcomp}
  \end{center}
\end{figure}

In order to demonstrate the reliability of the interferometric surface maps, we focus on the 2012 data.  We separated the data into two sets (2012 November 7, 21, 24; December 4, 7 and 2012 November 8, 22, 25; December 5, 8) and imaged them with SURFING (see Figure \ref{2012SURFcomp}). 
The two images clearly show the dark spot near the southern pole and the large spot structure in the mid-latitudes of the northern hemisphere.  The equatorial spots are less easily compared between the two images.  The difference in features is likely the result of the two data sets being nearly equal in phase, but not in data quality (2012 December 8 is a particularly poor-quality data set).  In combining the data sets for Figure \ref{Aitoff2012}, we obtain the highest quality image possible to date for $\sigma$~Gem.

\section{C.  Shifting the spectra to a common zero point}

In INVERS7PD, no correction for the orbital motion is included, and therefore the binary orbit has to be removed from the spectra before the inversions. In principle, two methods can be used for this task. The spectra can be cross-correlated and shifted to a common zero point, or the orbit of $\sigma$~Gem can be used to calculate the shifts needed for each phase. The disadvantage of using cross-correlation is that spots on the primary will affect the result, and can introduce wrong shifts between the spectra obtained at different phases with different spot configurations. On the other hand, for using the orbit in shifting the spectra, the orbit has to be very accurate.

\begin{figure}
 \centering
\begin{subfigure}
\centering
\includegraphics[angle=90,scale=.3]{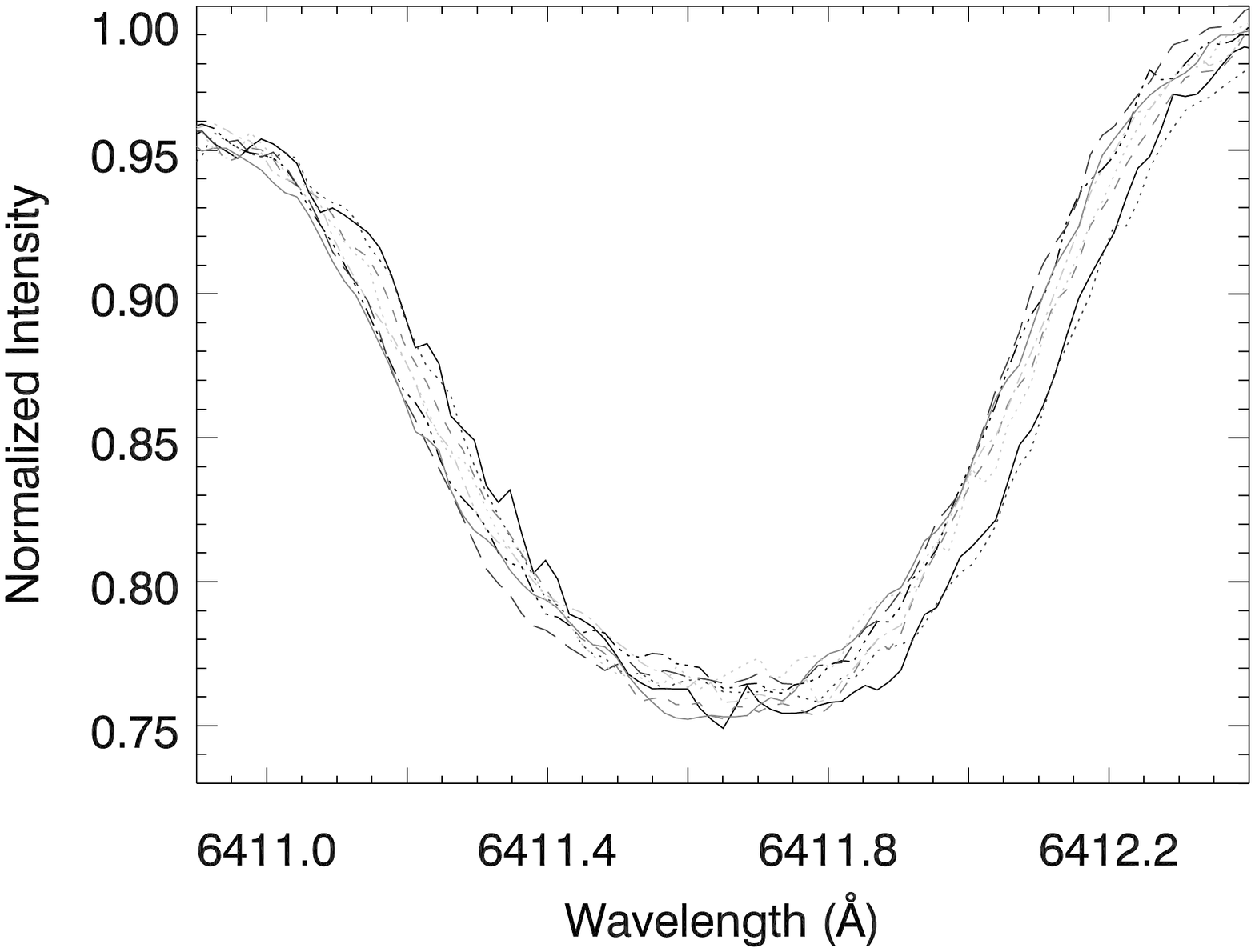}
\end{subfigure}
\begin{subfigure}
\centering
\includegraphics[angle=90,scale=.3]{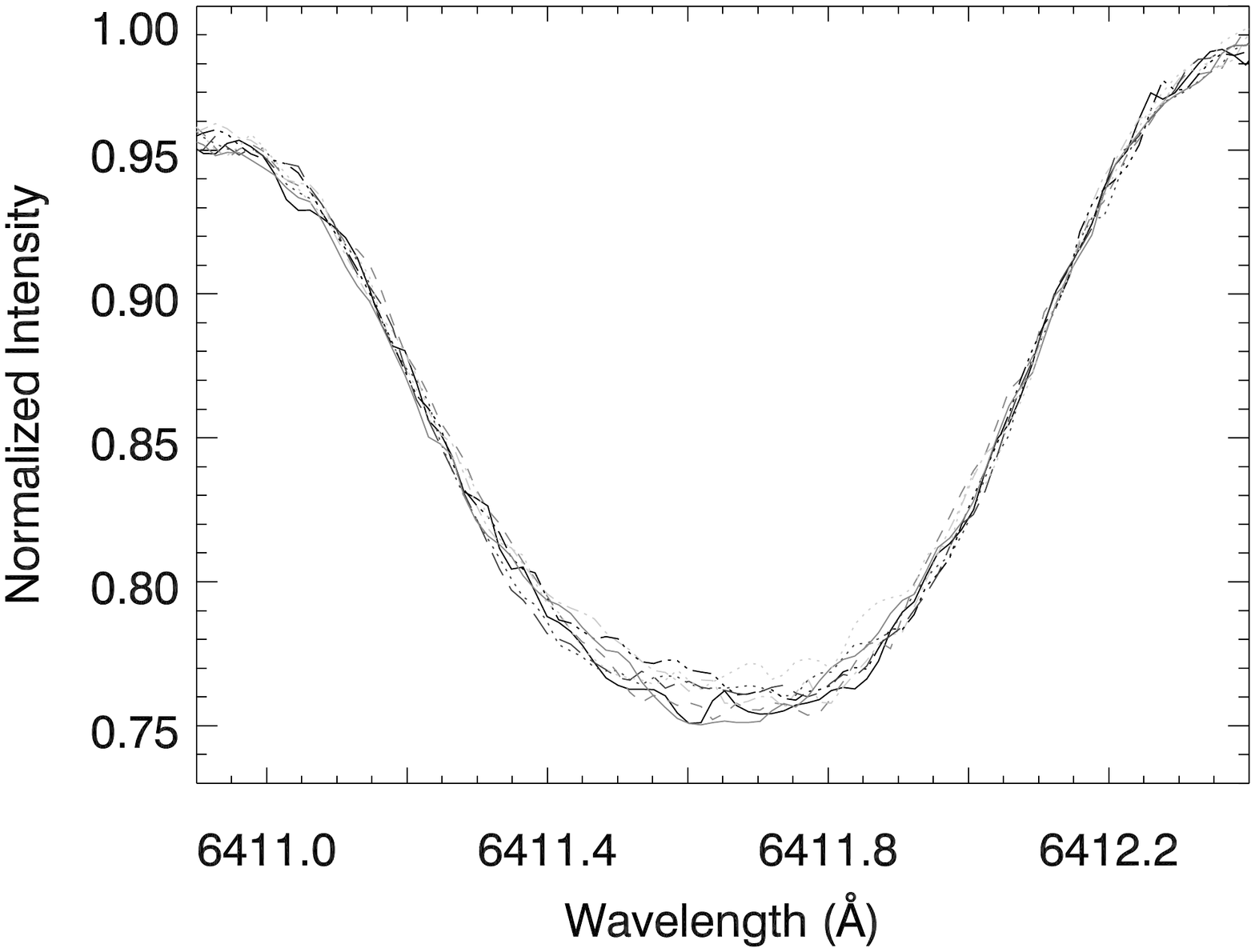}
 \end{subfigure}
  \caption{All Fe I 6411 line profiles of the UVES November 2012 data. Left: Line profiles shifted using the orbit by \citet{roe15}. Different line-styles denote different phases. Right: The same as on the left, but now the spectral lines profiles have been shifted using cross-correlation.}
  \label{lines2012}
\end{figure}

The orbit of $\sigma$~Gem has been recently determined by \citet{roe15} using radial velocity measurements of the primary and secondary and from the direct detection of the secondary with optical interferometry. This accurate orbit was used for shifting the $\sigma$~Gem spectra to the common zero point.  When over-plotting the spectra those shifted with the orbital solution show much more scatter in their exact positions than those shifted using cross-correlation (see Figure \ref{lines2012}).

\begin{figure}
 \vspace{-1cm}
 \centering
  \includegraphics[angle=0,scale=.35]{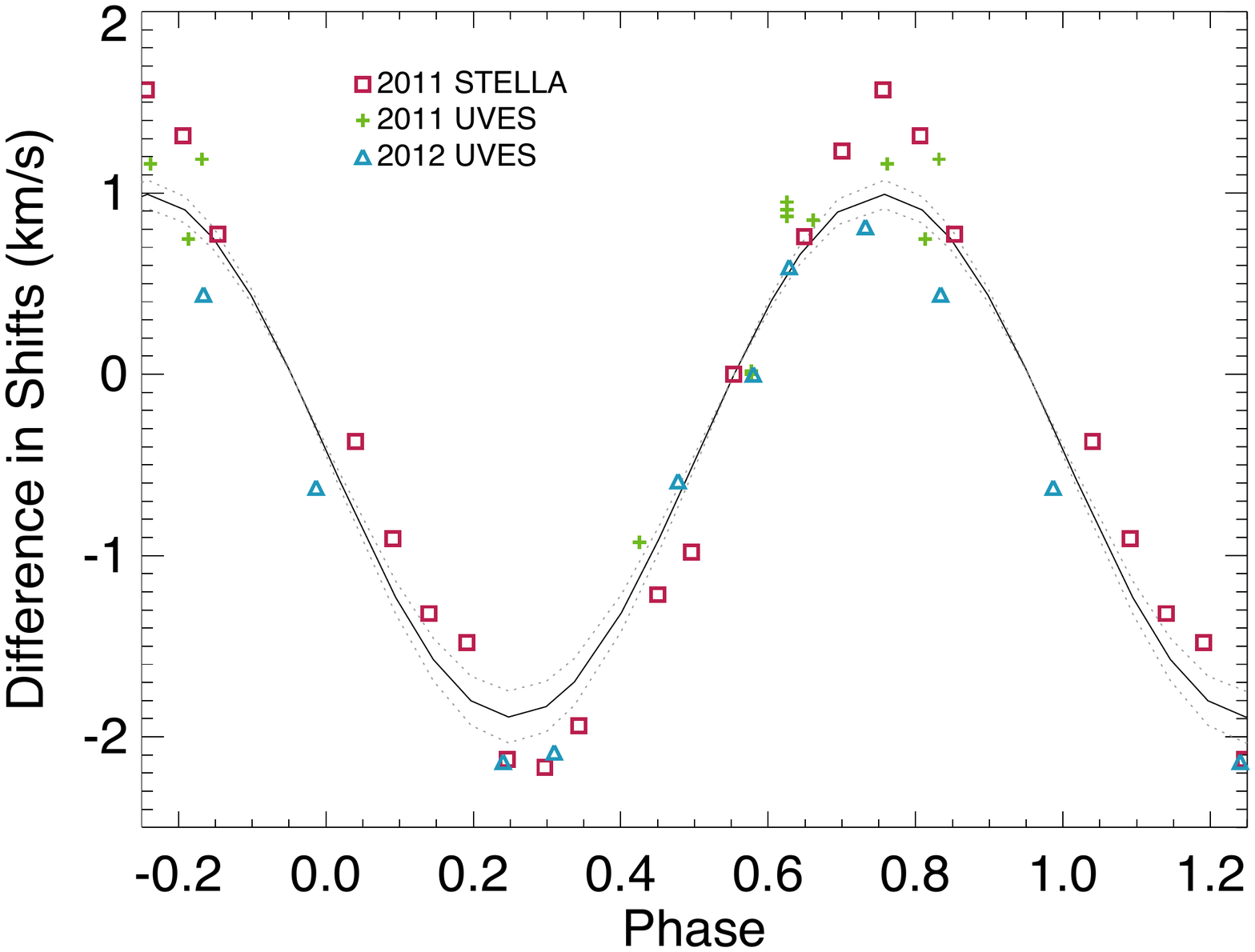} 
  \vspace{-0.5cm}
  \caption{Difference in shifts when cross-correlation results have been subtracted from the calculated 
 orbital shifts plotted against orbital phase. The symbols denote different data sets: red squares for STELLA 2011 data, green plus marks for UVES 2011 data, and blue triangles for UVES 2012 data. The black solid line indicates shifts obtained when $T_0$ is shifted by $+0.13$ days, and the gray dotted lines are 0.01-day uncertainties.  }
  \label{shifts}
\end{figure}

To investigate this discrepancy further, we take the difference between the shifts obtained with cross-correlation and orbital solution.  The cross-correlation only gives relative shifts, so the results are moved to the same zero-point as the orbital solution. This is done by shifting the observation closest to phase 0.5 to zero. As can be seen in Figure \ref{shifts}, 
the shifts are the same around phases 0.5 and 1.0. However, around the phases 0.2--0.4 the shifts obtained from cross-correlation are larger than the ones obtained from the orbit, and for the phases 0.6--0.8 the shifts from the orbit are larger. This same behavior is seen for both the STELLA and UVES data, and for both years of observation. This discrepancy can be explained if the orbit is shifted by 0.007 in phase, translating into a 0.13 day difference in the $T_0$ (time of nodal passage). For testing this, we have plotted the difference between $T_0$ \citep{roe15} and $T_0 + 0.13$ days (solid line in Figure \ref{shifts}). This modification in the $T_0$ reproduces the different results when using the cross-correlation and the orbit for shifting the spectra. Therefore, we have used $HJD_0 = 2453583.98 + 0.13 = 2453584.11$ for calculating the orbital velocities for moving the spectra. Obtained velocities are also given in Table \ref{spectra_log}.

We note that this 0.13 day change in $T_0$ \emph{should not} be taken as a new value for an accurate orbit. Most likely this difference results from combined effects from both the $T_0$ and the period. For the purpose of shifting the spectra to a common zero point for Doppler imaging the changed $T_0$ is accurate enough approximation, and therefore it is used here.

It is also worth noting that there are small shifts for each individual spectral line. These individual shifts are accounted for by doing inversions for each line separately, shifting the spectra, and finding the best-fit between the model and observations. These individual shifts are typically of the order of $0.01$ \AA.


\end{document}